\documentclass[review,authoryear,12pt]{elsarticle}

 \usepackage{graphicx}
 \usepackage{amsmath}
\usepackage{amssymb}
\usepackage{mathrsfs}
\usepackage{color}
 \usepackage{multirow}

\newcommand{\degrees}{\ensuremath{^\circ}}

\journal{Ultrasound in Medicine and Biology}

\begin{document}
\begin{frontmatter}

\title{Monodisperse versus polydisperse ultrasound contrast agents: nonlinear response, sensitivity, and deep tissue imaging potential}

\author[Affil1]{Tim Segers \corref{cor1}}
\author[Affil2]{Pieter Kruizinga}
\author[Affil1]{Maarten P. Kok}
\author[Affil1]{Guillaume Lajoinie}
\author[Affil2,Affil3]{Nico de Jong}
\author[Affil1]{Michel Versluis}

\address[Affil1]{Physics of Fluids Group, University of Twente, P.O. Box 217, 7500AE, Enschede
The Netherlands}
\address[Affil2]{Biomedical Engineering, Thoraxcenter, Erasmus MC, P.O. Box 2040, 3000 CA Rotterdam, The Netherlands}
\address[Affil3]{Acoustical Wavefield imaging, Delft University of Technology, P.O. Box 5, 2600 AA Delft
The Netherlands}

\cortext[cor1]{Corresponding Author: Tim Segers, Drienerlolaan 5, 7522NB, Enschede, The Netherlands; timsegers1@hotmail.com}

\begin{abstract}
Monodisperse microbubble ultrasound contrast agents have been proposed to further increase the signal-to-noise-ratio of contrast enhanced ultrasound imaging. Here, the sensitivity of a polydisperse preclinical agent was  compared experimentally to that of its size- and acoustically-sorted derivatives by using narrowband pressure- and frequency-dependent scattering and attenuation measurements. The sorted monodisperse agents showed up to a two orders of magnitude increase in sensitivity, i.e. in the average scattering cross-section per bubble. Moreover, we demonstrate here, for the first time, that the highly nonlinear response of acoustically sorted microbubbles can be exploited to confine scattering and attenuation to the focal region of ultrasound fields used in clinical imaging. This property is a result of minimal prefocal scattering and attenuation and can be used to minimize shadowing effects in deep tissue imaging.  Moreover, it potentially allows for more localized therapy using microbubbles through the spatial control of resonant microbubble oscillations.

\end{abstract}


\end{frontmatter}
\pagebreak

\section*{Introduction} \label{intro}

Ultrasound contrast agents (UCA) consist of a suspension of microbubbles that are stabilized against dissolution and coalescence by a surfactant shell, typically composed of biocompatible phospholipids. The compressibility of the microbubbles gas core allows for ultrasound driven radial bubble oscillations. 
The resulting nonlinear echo  can be used to visualize and quantify organ perfusion~\citep{Lindner2004}. Even though the scattering cross-section of of UCA microbubbles is typically 9~orders of magnitude higher than that of particles of the same size~\citep{deJong1991}, the scattering efficiency is quite low. 
The larger part of the incident acoustic energy is lost due to viscous damping.  The intermolecular viscous dissipation within the lipid shell accounts for approximately 80\% of the energy loss, the remainder is dissipated due the viscosity of the surrounding fluid and, in addition, due to thermal diffusion~\citep{Khismatullin2002, vanderMeer2007}. The energy loss results in the attenuation of an ultrasound wave propagating through a microbubble suspension~\citep{Leighton1994, deJong1992}. 

The radial microbubble oscillation amplitude in response to a driving ultrasound field is strongly dependent on the coupling between the frequency of the ultrasound field and the resonance frequency of the microbubble. The microbubble resonance frequency is inversely proportional to its size through the Minneart eigenfrequency~\citep{Minnaert1933}. On top of that, it is highly affected by the physical properties of the microbubble shell that can be modelled as a viscoelastic membrane with a shell viscosity, resulting in an increased damping, and with a shell elasticity, that increases the resonance frequency~\citep{vanderMeer2007}. Commercial UCA are available as a suspension of microbubbles with a relatively wide size distribution with radii typically ranging from 0.5~to~8~$\mu$m. 
Clinical ultrasound scanners operate over a relatively narrow frequency bandwidth, with respect to that of the resonance frequencies of the microbubbles present in a typical UCA. Thus, it is expected that only a small fraction of the UCA population attributes to the overall echo. Therefore, the sensitivity of contrast-enhanced ultrasound imaging, and more particularly that of single bubble molecular imaging~\citep{Klibanov2006}, can be substantially increased through the use of monodisperse bubbles that are resonant to the driving ultrasound pulse. 

The sensitivity increase that may result from the use of a monodisperse UCA was already suggested before~\citep{Talu2007, Hettiarachchi2007, Shih2013, Parrales2014, Kaya2010, Gong2014, Stride2009c, Segers2016d}. \emph{In-vitro} experiments have shown that the echoes of monodisperse bubbles are more correlated than that of a polydisperse population~\citep{Talu2007}. \emph{In-vivo} experiments in rats have shown a higher video intensity for monodispere bubbles as compared to a polydisperse agent~\citep{Streeter2010}. 

The potentially higher sensitivity of a monodisperse contrast agent was reported to be of main interest for molecular imaging~\citep{Klibanov2006} and drug delivery applications~\citep{Tsutsui2004, Hernot2008, Deelman2010, Carson2012, Dewitte2015} where typically only a small amount of bubbles is retained at the target site~\citep{Talu2007}. For blood pool imaging in humans, large amounts of microbubbles can be injected (on the order of one billion bubbles) to compensate for the lower sensitivity of a polydisperse agent. Therefore, monodispersity was thought to be of less importance here~\citep{Talu2007, Kaya2010}. However, it has been shown that the resonance behavior of narrow size distribution bubble populations is more narrowband, and more nonlinear, than that of a polydisperse agent~\citep{Emmer2009}. The strong driving pressure-dependent resonance behavior in particular~\citep{Overvelde2010, Xia2015, Segers2016d}, may result in a very different scattering behavior of a monodisperse agent as compared to that of a polydisperse agent in a typical ultrasound field employed for clinical contrast enhanced ultrasound imaging. The clinically used ultrasound beams are  focussed, with pressure amplitudes increasing towards the acoustic focal region, and deceasing thereafter~\citep{Segers2016d, Sojahrood2015}, resulting in the insonation of the UCA at a broad range of acoustic pressures.  A systematic experimental comparison between a polydisperse agent and a monodisperse agent with the same microbubble coating properties has never been conducted, neither to study sensitivity, nor to study the pressure-dependent scattering in a clinically relevant focused ultrasound field. 

A monodisperse microbubble suspension can be synthesized in a microfluidic flow-focusing device~\citep{Ganan-calvo2001a, Anna2003, Garstecki2005b,Segers2016}. Recently, the full parameter space for stable lipid-coated microbubble synthesis was characterized~\citep{Segers2017}. Alternatively,    a narrow size distribution bubble population can be obtained by sorting 
a polydisperse UCA, e.g., by means of filtration~\citep{Emmer2009}, decantation~\citep{Goertz2007d}, and centrifugation methods~\citep{Feshitan2009}. Microbubbles can be sorted with a higher degree of control in microfluidic devices, e.g., they can be sorted to size in a pinched microchannel~\citep{Kok2014} and they can be sorted to their resonance behaviour using the primary radiation force induced by a traveling acoustic wave~\citep{Segers2014}.  An advantage of sorting methods over the flow-focusing method is that sorting methods may allow for a direct comparison of the effects of the bubble size distribution on the acoustic properties of the polydisperse agent and its monodisperse derivatives, since the different populations originate from the very same native bubble population.

\begin{figure*}[ht] 
\begin{center}
\includegraphics[width=.951\textwidth]{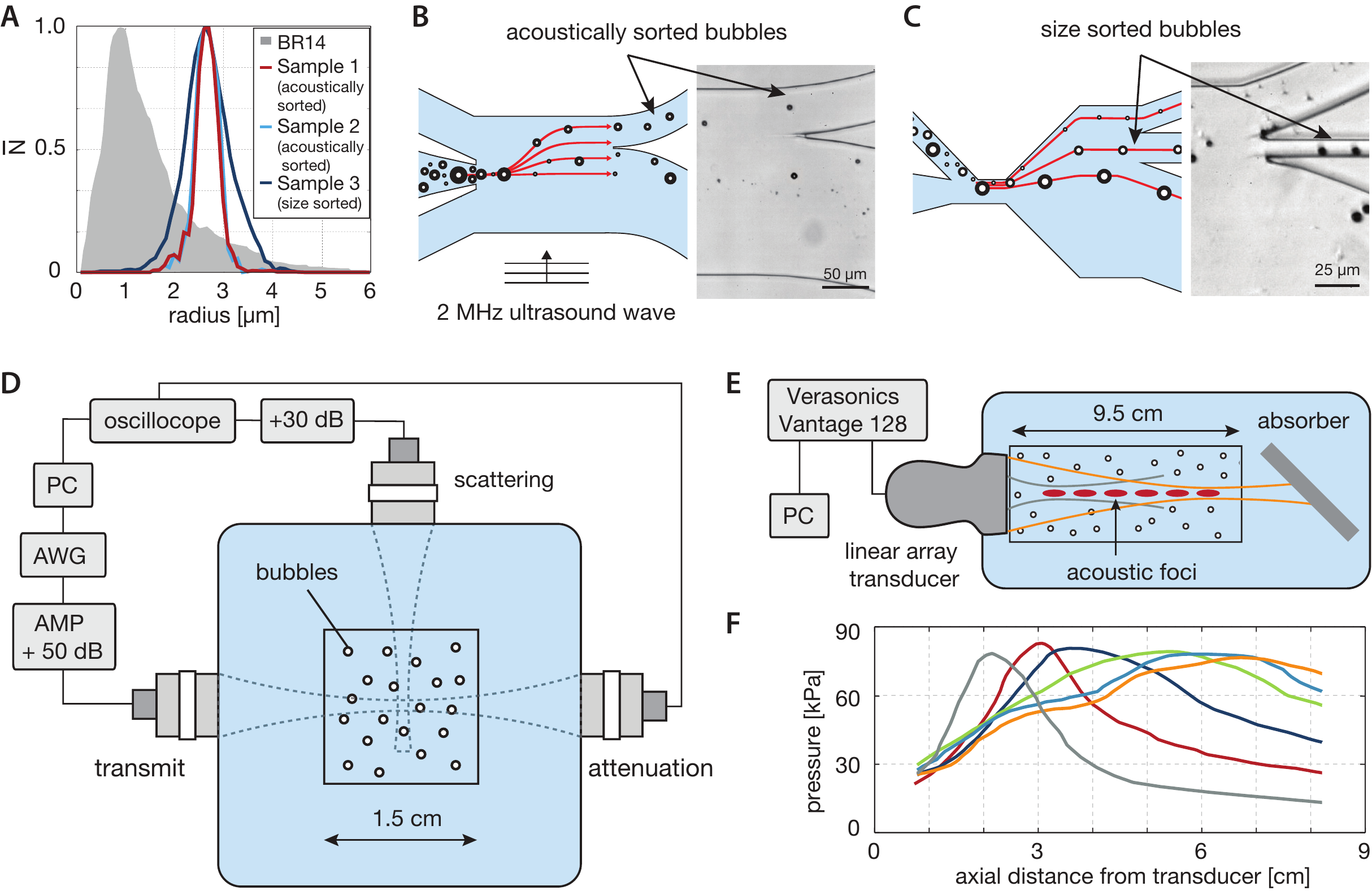}
\caption{(A) Size distribution of the native polydisperse contrast agent and of the sorted microbubble populations.  The narrow size distribution bubble populations were obtained through microfluidic sorting of the native agent in (B) an acoustic bubble sorting chip, where resonant bubbles are separated from non-resonant bubbles by the primary radiation force (Sample 1 and 2), and  (C) in a pinched flow fractionation (PFF) chip, where bubbles are sorted to size (Sample 3). (D) Schematic drawing of the acoustic characterization setup used to characterize the bubble populations by narrowband scattering and attenuation measurements. (E) Schematic drawing of the pulse-echo setup with a linear array transducer. The setup is used to study the nonlinear microbubble response in a typical focused ultrasound field employed for contrast-enhanced ultrasound imaging.  (F) The focal distances were dynamically varied from~2 to 7~cm in steps of 1~cm while the focal pressure was kept constant. }
\label{F:1}
\end{center}
\end{figure*}

The aim of this work was to characterize and to compare the nonlinear behavior and the sensitivity of a native agent  to that of its microfluidically sorted derivatives using pressure- and frequency dependent scattering and attenuation measurements. The systematic characterization was used to understand the pulse-echo response of the different bubble populations in a clinically relevant focussed ultrasound field of which the focal position and the focal pressure were varied. 

\section*{Materials and Methods}
\subsection*{Agent handling and bubble sorting procedures}

A polydisperse preclinical perfluorobutane-based ultrasound contrast agent (Bracco BR-14, Bracco Research Geneva) was used, containing bubbles coated with  
DSPC and DPPG lipids~\citep{Sijl2010}. It was reconstituted with  5~mL of Milli-Q water (Millipore Corporation, Billerica, MA, USA) and put to rest for at least 10~min to allow the bubbles to stabilize. The optically measured size distribution is shown in Fig.~\ref{F:1}A with a total bubble concentration of $2.5\times 10^8$~bubbles/mL. 

The native BR-14 agent was sorted to its acoustic property by using the acoustic bubble sorting method outlined by Segers~\emph{et al.}~\citep{Segers2014, Segers2016d}. In total, two acoustically sorted bubble populations were produced; Sample~1 and Sample~2. The sorting chip, shown in  Fig.~\ref{F:1}B, had an overall channel height of 14~$\mu$m. It comprised two outlet channels to separate the resonant from the non-resonant bubbles through the primary radiation force induced by a 2~MHz traveling acoustic wave.  The cross section of the sorting channel was 14~$\times$~ 200~$\mu$m$^2$ with a total length of 5~mm. The width of the resonant-bubble outlet was 50~$\mu$m (Fig.~\ref{F:1}B). The traveling wave was generated by a 6~mm diameter piezo driven by a  continuous-wave sinusoid with a 1.8~V amplitude (Tabor Electronics, WW1072, Tel Hanan, Israel).  The maximum acoustic pressure amplitude within the sorting channel was measured as described by~\cite{Segers2014} and it was 20~kPa.
The native BR-14 agent was diluted  1.5~times before it was infused in the sorting chip at rates of 1.0~$\mu$L/min and 2.0~$\mu$L/min, for Sample 1 and Sample 2, respectively.   The co-flow rate was always 25~$\mu$L/min.  The resulting pressure drop over the sorting channel was approximately 50~kPa~\citep{BruusTheoM}. The flow-rates were controlled by high-precision syringe pumps (Harvard Apparatus, PHD 2000, Holliston, MA).  Sample~1 was collected by running the sorting procedure for 25~min and Sample~2 was collected for 30 minutes. During the sorting procedures, a high-speed recording was captured every 3~min at a frame-rate of 5000~frames/s for the duration of 1~s. The high-speed recordings were analyzed in MATLAB to determine the size distribution of the sorted bubbles, see Fig.~\ref{F:1}A. The bubble size was measured from the inflection point of the optical intensity profiles of the individual bubbles~\citep{Segers2014}. The maximum of both size distributions is positioned at a bubble radius of 2.7~$\mu$m and both size distributions have a polydisperity index (PDI, ratio of the standard deviation to the mean radius) of approximately 9\%. The PDI of the size distribution of the native agent was 60\%. The high-speed recordings were also used to estimate the bubble concentration in the sorted samples through interpolation of the number of sorted bubbles.

The native BR-14 agent was also sorted to size in a pinched flow fractionation (PFF) lab-on-a-chip device to produce Sample~3. The sorting chip, see Fig.~\ref{F:1}C, had a channel height of 14~$\mu$m and the outlet channel was 10~$\mu$m in width. The sorting behavior of the applied PFF chip was fully characterized by Kok \emph{et al.}~\citep{Kok2014}. The agent handling and infusion into the PFF chip were as described for the acoustically sorted samples. Sample~3 was collected by running the PFF chip for 1~hour at an UCA flow-rate of 0.2~$\mu$L/min and at a co-flow rate of 12~$\mu$L/min. The measured size distribution is shown in Fig.~\ref{F:1}A. 

Narrowband scattering and attenuation measurements were performed on an acoustically sorted bubble suspension (Sample~1), on a size-sorted bubble suspension (Sample~3), and on the native BR-14 agent. Before the characterization experiments, the sorted samples were diluted with water (Milli-Q, Millipore Corporation, Billerica, MA, USA) in a vial to a total volume of  22.5~mL after which the bubble concentration of Sample~1 was $4.0\times 10^3$~bubbles/mL and that of Sample~3  $5.8\times 10^3$~bubbles/mL. The native agent was diluted by 15.000~times to a bubble concentration of $1.7\times 10^4$~bubbles/mL. Aliquots of 4.5~mL were acoustically characterized.

Pulse-echo measurements were performed using a phased array transducer on acoustically sorted bubble Sample~2 and on the native BR-14 agent. To this end, Sample~2 was diluted in water (Milli-Q, Millipore Corporation, Billerica, MA, USA) to a total volume of 50~mL with a bubble concentration of $4.1\times 10^4$~bubbles/mL. The native agent was diluted by 3.000~times to a bubble concentration of $8.3\times 10^4$~bubbles/mL.

\subsection*{Narrowband scattering and attenuation measurements}
Scattering and attenuation spectra were measured simultaneously in a water tank at room temperature using a fully automated procedure that was controlled from a PC as described in detail by~\cite{Segers2016d}, see Fig.~\ref{F:1}D. In short, the bubble suspension, confined in a $1.5 \times 1.5 \times 4.5$~cm$^3$ acoustically transparent container, was insonified by a series of 100~narrowband 16-cycle ultrasound pulses per transmit frequency over a frequency range from 0.7~to~5.5~MHz in steps of 0.1~MHz at a constant acoustic pressure. The peak negative acoustic pressures employed during this study were 10~kPa, 25~kPa, 50~kPa, and 100~kPa. The transmit signals were generated by an arbitrary waveform generator (8026, Tabor Electronics) connected to an 50~dB amplifier (350L, E\&I, Rochester, NY). The beam of the transmit transducer (A305S, Panametrics-NDT, 2.25~MHz, 19~mm aperture, 25.4~mm focal distance) was focused in the center of the sample container and confocally aligned to a second transducer (C308, Panametrics-NDT, 5~MHz, 19~mm aperture, 25.4~mm focal distance) placed on the same center axis to measure attenuation. The received attenuated signals were processed offline in Matlab as follows: first they were gated around the transmit pulse-length, a power spectrum was calculated for each signal, the 100~power spectra per $f_T$ were averaged, and the attenuation at $f_T$ was calculated from the amplitude of the power spectrum at $f_T$ obtained from a measurement with bubbles present $|V_{bub}(f_T)|^2$ and from that without bubbles present $|V_{ref}(f_T)|^2$, as follows:
\begin{equation}\label{Eq:att}
\alpha = \frac{10}{d} \log\left( \frac{| V_{bub}(f_T)|^2}{|V_{ref}(f_T)|^2} \right),
\end{equation}
where $d$ is the acoustic path length over which the bubbles were present. 

Scattering was measured by a third transducer (Vermon, SR 885C1001, 3~MHz, 2$^{\prime \prime}$, 1$^{\prime \prime}$ aperture) positioned under a 90\degrees angle with respect to the transmit beam with its focus confocally aligned with that of the transmit beam.
The scattering signals were amplified by 30~dB (5077PR, Panametrics-NDT) and recorded by an oscilloscope (TDS5034B, Tektronix, Beaverton, OR) operated in sequence acquisition mode at a 50~MHz sampling rate. The signals were gated as before, a power spectrum was calculated for each signal, it was averaged over the 100~pulses per $f_T$ and the scattering coefficient at $f_T$ was calculated as follows: 
\begin{equation} \label{F:sc}
\eta = \frac{|V_{scat}(f)|^2}{|V_{ref}(f_{T})|^2} \frac{16z^2}{D^2},
\end{equation}
where $|V_{scat}(f)|^2$ is the power spectrum of the recorded scattering signal and  $|V_{ref}(f_T)|^2$  is the total transmitted power at the transmit frequency $f_T$ measured from the reflection of the transmit pulse from a stainless steel reflector placed at a  45\degrees angle at the position of the acoustic focus. The fundamental scattering coefficient $\eta_{fun}$ at $f_T$ is the magnitude of $\eta$ at $f_T$ and the scattering coefficient at the second harmonic $\eta_{2H}$ is the magnitude of $\eta$ at two times $f_T$. The scattering coefficient was compensated for the limited aperture $D$ of the receiving transducer with focal distance  $z$~\citep{DeJong1993}.  

\subsection*{Pulse-echo measurements using a linear array transducer}
 \subsubsection*{Experiments}
The pulse-echo experiments were conducted in a water tank at room temperature using a 96~element phased array transducer (P4-1 ATL) connected to a research ultrasound system (Vantage 256, Verasonics), see Fig.~\ref{F:1}E. 
The bubble suspension was confined in a $3.5\times 9.5\times 1.5$~cm$^3$ acoustically transparent container fabricated from polystyrene membrane exactly as described by~\cite{Segers2016d} with one side in contact with the ultrasound probe. 
The axial focus of the ultrasound field was positioned at 2, 3, 4, 5, 6, and 7~cm through beamforming of the transmit signals. The transmit aperture was apodized using a Hamming window such that the f-number $f_\#$ (ratio of the focal depth to the aperture width) was constant at $f_\#=2$. Thus, the width of the acoustic focus, and lateral resolution, was the same for all focal depths. The bubbles were insonified  20~times per focal depth by an 8-cycle ultrasound pulse at an ultrasound frequency of 1.5~MHz. The ultrasound frequency of 1.5~MHz was chosen since it corresponds to the frequency of maximum response $f_{MR}$ of the all bubble samples at the higher driving pressures. The transmit pulses were tapered by a Gaussian envelope over 1~cycle on each side of the pulse. The acoustic transmit pressure at the centerline of the probe was measured in pure water using a 0.2~mm hydrophone (Precision Acoustics, Dorset, UK), see Fig.~\ref{F:1}F. The transmit focal pressures were constant over the different focal depths. Scattering was measured at a sampling rate of 10~Ms/s for focal peak-negative pressures of 20~kPa, 30~kPa, 40~kPa, 60~kPa, and 80~kPa.  

The received signals of all transducer elements were stored and processed off-line using MATLAB. First, the signals were beam-formed and apodized using a synthetic aperture that was equal to that of the transmit aperture. After summation, the resulting scan line was divided in 46~signals of equal length corresponding to the depth range of 0~to~8.5~cm in steps of 1.9~mm. For each signal, a power spectrum was calculated and the scattered power at the fundamental was taken as the amplitude of the power spectrum averaged over a 100~kHz band around a frequency of 1.5~MHz. Finally, the scattered power at the second harmonic was that averaged over a 200~kHz band around a  frequency of 3~MHz.

\subsubsection*{Modeling the pulse-echo response}
From the measured pressure-dependent attenuation, the scattered power $|P_s|^2$ of a suspension of monodisperse bubble with radii $R_0$, at concentration $N$, can be calculated since they are directly proportional, as follows~\citep{Goertz2007d, Segers2016d}: 
\begin{equation}
\frac{|P_S|^2}{ |P_T|^2} = c \alpha
\end{equation}
where $|P_A|^2$ the power of the incident wave and $c= \delta_{rad} \delta_{tot}^{-1} \ln(10)(40 \pi N R_0^2)^{-1}$. The proportionality factor $c$ is constant for acoustically sorted bubbles due to their uniform shell parameters~\citep{Segers2016d}, i.e. the total damping $\delta_{tot}$ and the radiation damping $\delta_{rad}$ are equal for all bubbles within the suspension. Thus, the normalized scattered power at a given location $n$ in the transmit field is given by $c\alpha(P_{A,n})$, with $P_{A,n}$ the local pressure amplitude. However,  scattering from bubbles in between the transducer aperture and location $n$ attenuates the pressure amplitude, in turn lowering the scattering at location $n$. Therefore, the transmit-pressure amplitude, $P_A$ in Fig.~\ref{F:1}F, was corrected for attenuation by forward integration over the discretized imaging depth, as follows:
\begin{equation}\label{Eq:model}
P_{C,n} = \begin{cases} P_{A,1}, & \mbox{if } n = 1 \\   P_{A,n} 10^{-\Gamma_{n-1}/20}, & \mbox{if } n > 1 \end{cases}
\end{equation}
where $P_{C,n}$ is the pressure amplitude corrected for attenuation at grid point $n$. The total attenuation $\Gamma_n$ of the transmit wave up to grid point $n$ is given by:
\begin{equation}\label{Eq:model2}
\Gamma_n = \sum_{n=1}^{n} \alpha(P_{C,n})dx
\end{equation}
where $\alpha(P_{C,n})$ is the attenuation at the corrected pressure amplitude $P_{C,n}$, and $dx$ is the grid spacing. 

\section*{Results} 
\subsection*{Narrowband scattering and attenuation measurements}
\begin{figure*}[htbp] 
\begin{center}
\includegraphics[width=.9\textwidth]{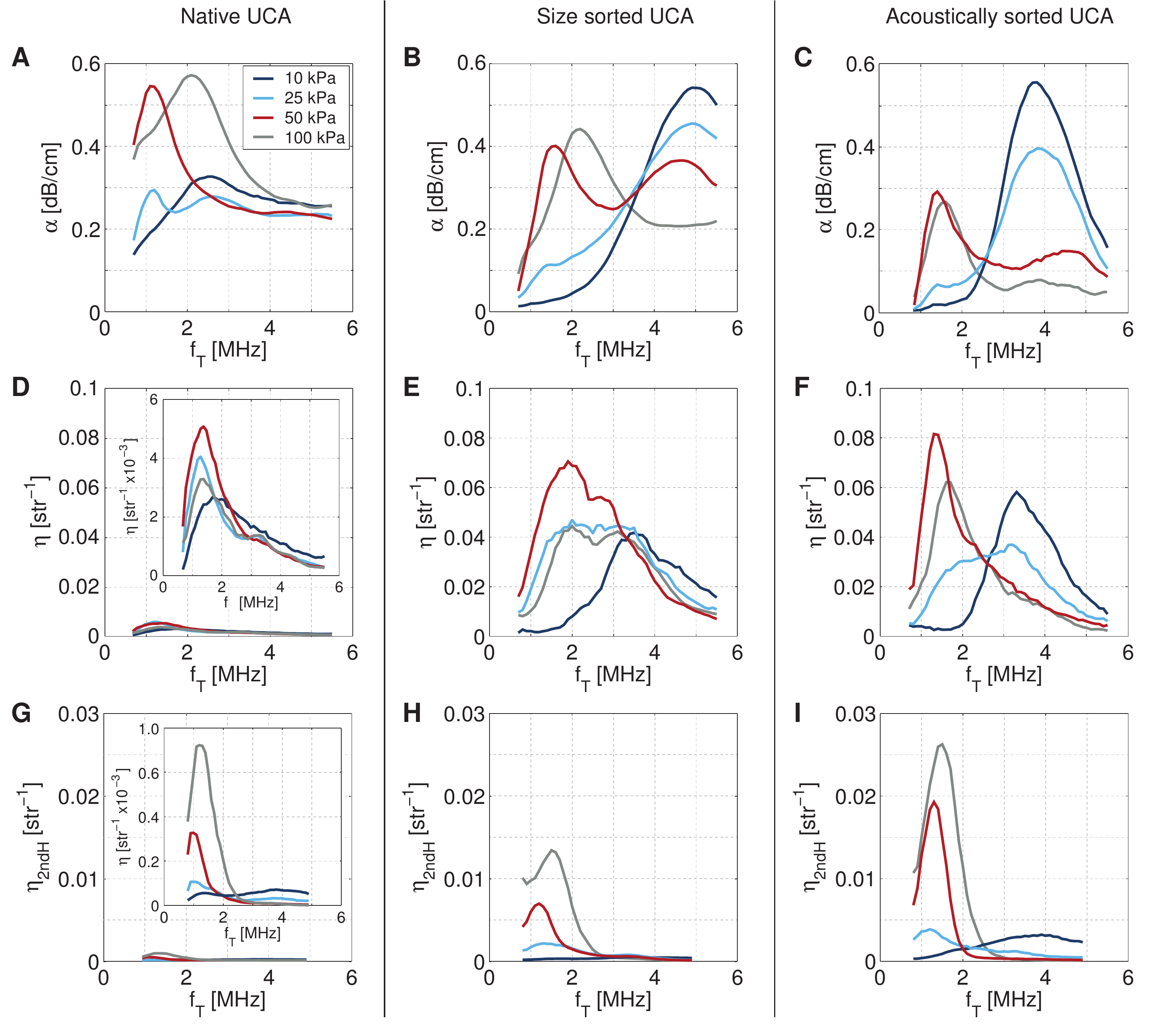}
\caption{Measured frequency dependent attenuation coefficients of (A) the native contrast agent, (B) the size sorted agent, and (C) the acoustically sorted agent. Attenuation curves were measured at peak negative pressures of 10, 25, 50, and 100~kPa. The simultaneously measured scattering coefficients at the fundamental requency are shown in Figs.~D-F and the second harmonic scattering coefficients are shown in Figs.~G-I. }
\label{F:3}
\end{center}
\end{figure*}

The measured scattering and attenuation coefficients are shown in Fig.~\ref{F:3}. The first column presents the data of the native agent, the second column that of the size-sorted agent. The third column shows data of the acoustically sorted agent which was also presented by~\cite{Segers2016d}. The attenuation curves of the three agents are plotted in the first row, see Figs.~\ref{F:3}A-C. The typical pressure-dependent response of the bubble populations was found to be similar; the frequency of maximum response decreases for increasing acoustic pressures. However, a closer look reveals a clear difference between the behavior of the sorted and the native agent. At a pressure of 10~kPa the sorted agents present almost no attenuation at low insonation frequencies whereas the native agent has an offset of approximately 0.2~dB/cm in its attenuation curves, which is at least 30\% of its peak value. On top of that, the absolute attenuation at $f_{MR}$ of the native agent increases with increasing acoustic pressure while that of the sorted agents decreases with increasing pressures. The attenuation curves of the bubble populations at a pressure of 100~kPa have a frequency of maximum response that is higher than that at a pressure of 50~kPa which indicates a decrease in the mean bubble size. Thus, the high number of insonations at a driving pressure of 100~kPa changed the size distributions of the bubble suspensions.

The scattering coefficients at the fundamental frequency are shown in Figs.~\ref{F:3}D-F. The scattering property of the native agent is very different from that of the sorted bubble populations. First, the scattering coefficients are about 10~times lower while the absolute magnitude of the  attenuation spectra was very similar for all bubble suspensions. Second, the scattering curves of the sorted agents present a stronger pressure and frequency dependency, i.e., the difference between the frequency of maximum response at low and high acoustic pressure is 100\%, while that of the native agent is only 30\%. On top of that, at 50~kPa, the frequency dependency of the scattering coefficient of the acoustically sorted agent is very similar to that of the native agent while that at 10~kPa it is very different. Thus, bubble sorting increases the non-linear, pressure dependent, response of the contrast agent. The non-linear response is maximized for the acoustically sorted agent that presents a more narrowband scattering than the size-sorted agent at all insonation pressures. 

The scattering coefficients at the second harmonic of the transmit frequency are plotted in Figs.~\ref{F:3}G-I. Similar to the fundamental scattering coefficients, the second harmonic scattering coefficients of the sorted agents are at least 10 times larger than that of the native agent. In addition, the second harmonic scattering coefficient of the acoustically sorted agent is approximately 3 times larger than that of the size sorted agent and at 100~kPa it is even 30~times larger than the second harmonic scattering coefficient of the native agent. 

\begin{figure}[t] 
\begin{center}
\includegraphics[width=1\columnwidth]{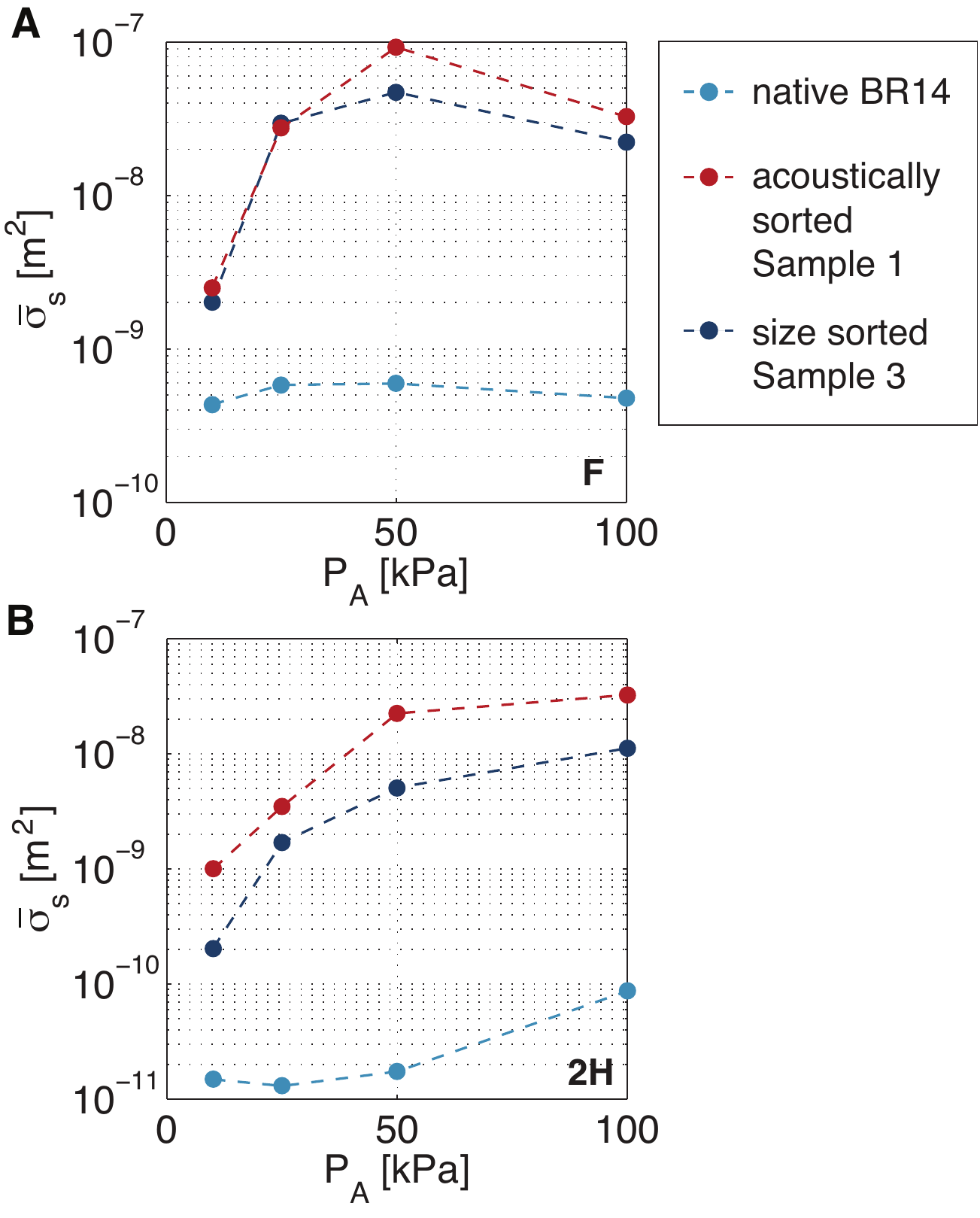}
\caption{ Sensitivity expressed as the mean scattering cross-section per bubble $\overline \sigma_s$, of the acoustically sorted agent, of the size sorted agent, and of the native BR-14 agent plotted at (A) the fundamental frequency and (B) at the second harmonic as a function of the insonation pressure. }
\label{F:star}
\end{center}
\end{figure}

\begin{figure*}[htb!] 
\begin{center}
\includegraphics[width=.9\textwidth]{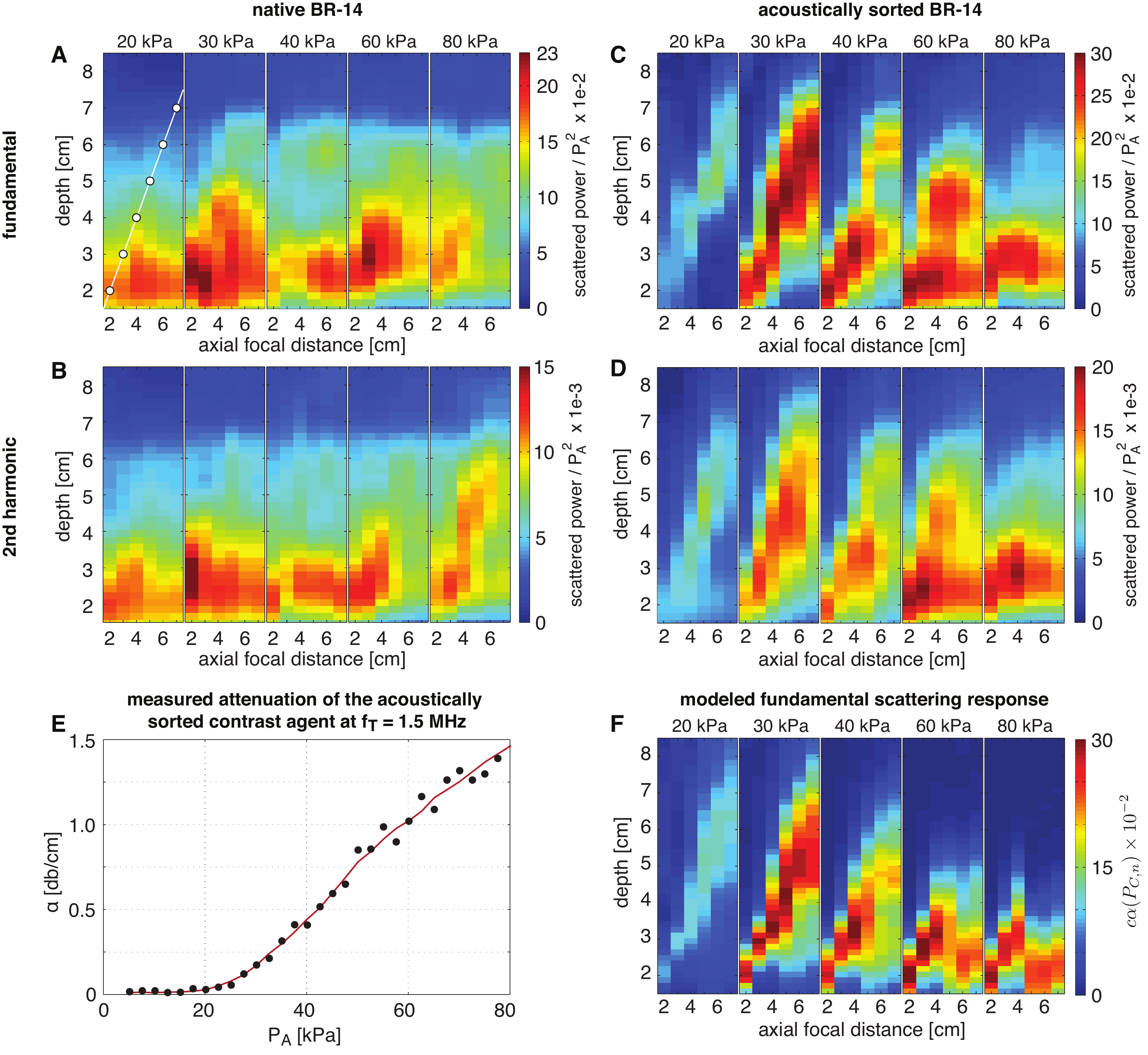}
\caption{ (A) Scattered power at an ultrasound frequency of 1.5~MHz normalized by the power of the transmit pulse as a function of the imaging depth for axial focal distances of 2, 3, 4, 5, 6, and 7~cm for the (A) native agent and (C) for the acoustically sorted agent.  The white dots in (A) indicate the focal regions of the transmitted acoustic pressure fields. (B) Scattering response at the second harmonic plotted for the native agent and (D) for the acoustically sorted agent. (E) Attenuation of the acoustically sorted  agent at an ultrasound frequency of 1.5~MHz as a function of the acoustic insonation pressure. (F) Modeled scattering response of acoustically sorted BR-14 microbubbles.}
\label{F:4}
\end{center}
\end{figure*}

The scattering response of all bubble populations is maximum at a frequency of approximately 1.5~MHz for driving pressures $P_A$ exceeding 20~kPa. Therefore, the sensitivity, or average scattering cross-section per bubble $\overline \sigma_s$, was calculated at a driving frequency of 1.5~MHz from the ratio of the corresponding scattering coefficient to the bubble concentration. The obtained sensitivity at the fundamental frequency is plotted in Fig.~\ref{F:star}A and that at the second harmonic  in Fig.~\ref{F:star}B, both as a function of the insonation pressure. For non-linear bubble oscillations, i.e. $P_A \geq 20~kPa$, the fundamental $\overline \sigma_s$ of the sorted agents is on average almost 2~orders of magnitude larger than that of the native agent. The  $\overline \sigma_s$ at the second harmonic is on average even 3~orders of magnitude larger than that of the native agent. Thus, to obtain equal scattering levels for the polydisperse and for the sorted agents, the polydisperse bubble concentration has to be 100~times higher. For molecular imaging, were only small amounts of bubbles are retained at the target site, this implies that at equal targeting concentrations, the scattered power of the sorted agents will be 2 orders of magnitude higher. 

\subsection*{Pulse-echo measurements using a phased array probe}
The measured scattered power normalized by the focal power of the transmit pulse ($P_A^2$) is plotted in Fig.~\ref{F:4} as a function of imaging depth and focal distance for the different acoustic pressure amplitudes. For the native BR-14 agent, the normalized scattered power at the fundamental frequency and that at the second harmonic are plotted in Figs.~\ref{F:4}A and \ref{F:4}B, respectively. The maxima of the normalized scattered power at the fundamental frequency, and at the second harmonic, are located within an axial distance of 3~cm from the transducer aperture, independent of the focal distance. The focal regions of the transmit pressure fields are indicated by the white dots in Fig.~\ref{F:4}A. Furthermore, no clear relationship is observed between the spatial scattering distribution and the acoustic focal pressure. 

For the acoustically sorted agent, the spatial distribution of the scattered power at the fundamental frequency and at the second harmonic is highly dependent on the focal pressure and on the focal distance of the acoustic field, see Fig.~\ref{F:4}C and D, respectively. At a focal pressure of 20~kPa, the maxima of the scattered power are spatially distributed around the focal regions (Fig.~\ref{F:1}F) of the ultrasound field, with near zero scattering before and after the acoustic focus. However, the maximum normalized scattered power is lower than that at higher driving pressures. At a focal pressure of 30~kPa, the maxima of the scattered power are also spatially distributed around the focal regions of the ultrasound field, but the maximum normalized scattered power amplitudes are now comparable to those obtained at higher focal pressure. A further increase in the acoustic focal pressure to 40~kPa and beyond results in a spatial shift of the scattered power maxima  towards the transducer aperture and thus, away from the high driving pressure region in the acoustic focus of the transmit field (Fig.~\ref{F:1}F). 

\section*{Discussion}
The highly focal-pressure dependent spatial scattering distribution observed for the acoustically sorted agent can be explained from its narrow-bandwidth, pressure dependent, resonance behavior that was fully characterized and presented in Fig.~\ref{F:3}. At acoustic pressures below 20~kPa, the acoustically sorted bubbles resonate at 3.5~MHz, and at acoustic pressure above 20~kPa they resonate at 1.5~MHz.  Thus, there is a critical pressure in the acoustic driving above which the resonance frequency starts to shift from the elasticity-dominated regime at a higher frequency to a  the elasticity-free regime at lower ultrasound frequency.  Therefore, at the employed insonation frequency of 1.5~MHz and at a focal pressure of 20~kPa, scattering is only observed from the acoustic focus of the transmit field where the acoustic pressure is just high enough to drive the bubbles into resonance. Increasing the focal pressure to 30~kPa increases the scattering efficiency, i.e. the normalized scattered power increases while the scattered energy is still confined to the focal region. By increasing the focal pressure even further, the acoustic pressure exceeds the critical pressure (20 kPa) for resonant microbubble oscillations in regions outside the acoustic focus resulting in strong scattering of the transmitted ultrasound waves before the acoustic focus is reached. As a consequence, the ultrasound wave is attenuated, i.e., its pressure amplitude decreases during its propagation towards the focal region and, accordingly the scattered microbubble echo amplitude that is directly proportional to the pressure amplitude of the ultrasound pulse~\citep{Church1995}. Thus, attenuation of the ultrasound wave  causes the observed maximum in scattered power to shift towards the transducer aperture at pressure amplitudes exceeding 40~kPa. This so-called 'shadowing' effect is frequently observed in contrast enhanced deep tissue imaging~\citep{Szabo2004}. However, here, it is shown that by carefully selecting the focal pressure, microbubble scattering can be confined to the focal region of the acoustic transmit field and that, thereby,  shadowing effects can be minimized.

To support the above explanation for the observed focal-pressure dependent spatial scattering distribution of the acoustically sorted agent, its scattering response was modelled using the high-precision pressure dependent attenuation measurements recently reported by~\cite{Segers2016d} and reproduced in Fig.~\ref{F:4}E for an ultrasound frequency of 1.5~MHz. 
The attenuation data plotted in Fig.~\ref{F:4}E was smoothed (solid curve) and directly used in Eq.~\ref{Eq:model} that was solved together with Eq.~\ref{Eq:model2} by forward integration in MATLAB using  a grid spacing of 0.5~mm. The modeled normalized scattered power $c\alpha(P_{C,n})$ of the acoustically sorted agent is plotted in Fig.~\ref{F:4}F as a function of the imaging depth and focal distance. Note that it is in good agreement with the measured scattering response in Fig.~\ref{F:4}C, which demonstrates that at focal pressures higher than 30~kPa, the measured decrease in echo amplitude at the deeper regions is, indeed, caused by attenuation of the acoustic transmit field due to resonant bubble oscillations. 

The spatially more uniform focal- and acoustic pressure independent scattering response of the native BR-14 agent can now also be explained. The narrowband scattering and attenuation coefficients plotted in the first column of Fig.~\ref{F:3} show that at all driving pressures and at an insonation frequency of 1.5~MHz, there is always a strong response from the native agent.  The strong response at all driving frequencies and acoustic pressures results from the broad microbubble size distribution as the bubble suspension always contains bubbles with a size resonant to the driving ultrasound pulse. Thus, in the pulse-echo experiment, at any focal pressure, there are always resonant microbubble oscillations attenuating the propagating ultrasound field resulting in a significant attenuation in the region between the transducer aperture and the focal region.  

The new insight presented in this work allows for the development of new imaging schemes to exploit the nonlinear response of monodisperse microbubbles in order to minimize shadowing effects in deep tissue imaging. The acoustic field can be designed such that the transmitted ultrasound wave 'tunnels' through the contrast agent, i.e. there is minimal scattering and attenuation, except for the focal region.  It is good to note that contrast-enhanced ultrasound imaging (CEUS) is typically performed at peak negative acoustic pressures on the order of 100 kPa or higher. However, the sensitivity of the sorted agents was found to be 2 to 3 orders of magnitude higher than that of the native agent. The higher sensitivity potentially allows for the use of lower acoustic pressures in CEUS, which are then optimal to exploit the tunneling effect. Furthermore, custom-made transducers can be developed with a high sensitivity, i.e. to be able to lower the transmit pressure, and with a high focal gain, e.g. through the use of a large aperture and where the lower prefocal pressure aids the tunneling effect. On top of that, the shell properties of monodisperse bubbles can be further optimized to increase their nonlinear response and the onset acoustic pressure for nonlinear oscillations.

It is also good to note that the spatial control over resonant microbubble oscillations increases with an increasing driving ultrasound pulse length.  This can be appreciated from Fig.~S1 in the supplementary information that shows the pulse-echo scattering response of the acoustically sorted agent for a 4-cycle, 1.5~MHz transmit pulse. The tunneling effect can be clearly appreciated for the shorter 4-cycle pulses, however, the scattering response is spatially distributed over a larger area.   Thus, the confinement of the scattering properties of the acoustically sorted agent is aided by narrowband transmission pulses, which is, apart from Doppler, not immediately ideal for high-resolution imaging. However, more advanced imaging schemes, such as for example proposed by~\cite{Meral2013} where spectrally randomized transmissions are used to build up an image, may provide an interesting alternative to this problem. Furthermore, techniques such as coded excitation in conjunction with pulse compression, may be exploited to regain resolution loss obtained with long imaging pulses~\citep{ODonnell1992,Misaridis2000, Song2015}.
On the other hand,  the spatial control over resonant microbubble oscillations using narrowband excitation pulses is ideally suited for therapeutic applications for the localized delivery of a microbubble payload, to locally induce microstreaming, and to locally palpate or sonoporate tissues~\citep{Cock2016,Helfield2016}.

The narrowband response of the sorted agents results directly from their narrow size distribution and their acoustic homogeneity~\citep{Segers2016d}. The stronger driving pressure-dependent response of the sorted agents is most likely also in part due to the narrow size distribution. On top of that, there may be a contribution resulting from a difference in shell stiffness with respect to that of the native BR-14 bubbles. The bubbles are sorted at an overpressure required to drive the flow through the sorting chip. The resulting decrease in microbubble surface area may increase shell stiffness due to the larger intermolecular forces between the more closely packed lipid molecules. Moreover, a decrease in microbubble surface area may overcompress the lipid shell leading to a selective loss of shell material through which shell properties may change~\citep{Segers2016}. However, until now this aspect on the differences in the nonlinear behavior of the different populations remains inconclusive and further research on the detailed properties of the lipid shell under increased ambient pressure, shear flow, and acoustic forcing is required.

It is of interest to say a few words on the repeatability of the experiments. The data of the acoustically sorted agent in Fig.~\ref{F:3}, Fig.~4A-D, Fig.~4E, and Fig.~S1 (supplementary information) was obtained on different days, using different BR-14 vials and using different sorted bubble populations. The spatial distribution of the scattering measured in the pulse-echo setup and shown in Fig.~4B can be accurately modeled using the high-precision attenuation data shown in Fig.~4E.  Moreover, Fig.~4B can be explained using Fig.~2. The same holds for the spatial scattering distribution of the native BR-14 agent measured in the pulse-echo setup, see Fig.~4A, which can be explained using the attenuation data of the native agent shown in Fig.~2A. Thus, all sorted bubble populations give identical results and demonstrate the exceptional repeatability of the experiments.

To exploit the tunneling effect during contrast-enhanced ultrasound imaging over large imaging depths, e.g. to detect multiple focal lesions in a 10~cm liver, an imaging scheme with focused beams at multiple focal depths could be developed. By combining the separate images into a new image the full liver can be visualized. 

For plane wave imaging~\citep{Tanter2014}, the tunneling effect can potentially be used by transmitting ultrasound at a frequency that matches the resonance frequency of the microbubbles at low acoustic driving pressures. A transmitted plane wave will then not be scattered and attenuated until its pressure amplitude has decreased, e.g. due to tissue attenuation and dispersion, below the threshold for linear bubble oscillations, i.e. below 20 kPa for the acoustically sorted bubbles.

\section*{Conclusions}
\label{Conclusions}
The sensitivity of size- and acoustically-sorted microbubble populations is typically 2~orders of magnitude higher than that of a polydisperse ultrasound contrast agent. At the second harmonic, it is even 3~orders of magnitude higher. The sensitivity of the acoustically sorted agent was typically 2~times higher than that of the size sorted bubble population. Moreover, the resonance behavior of sorted ultrasound contrast agents is more nonlinear than that of a polydisperse agent. The highly nonlinear response of acoustically sorted microbubbles can be exploited to minimize shadowing effects in deep tissue imaging since the resonant microbubble oscillations can be limited to the focal region of the ultrasound field, with near zero scattering and attenuation outside the focal region. 

\section*{Acknowledgements}
\label{Ack}
We thank Peter Frinking for stimulating discussions. 
We thank Bracco Research Geneva for the supply of BR-14 ultrasound contrast agents. We also want to thank Gert-Wim Bruggert, Martin Bos, and Bas Benschop for their skilful technical assistance.  This work is supported by NanoNextNL, a micro and nanotechnology consortium of the Government of the Netherlands and 130 partners.  

\pagebreak




\bibliographystyle{UMB-elsarticle-harvbib}


\pagebreak

\section*{Figure Captions}
\begin{description}
\item[Figure 1:] (A) Size distribution of the native polydisperse contrast agent and of the sorted microbubble populations.  The narrow size distribution bubble populations were obtained through microfluidic sorting of the native agent in (B) an acoustic bubble sorting chip, where resonant bubbles are separated from non-resonant bubbles by the primary radiation force (Sample 1 and 2), and  (C) in a pinched flow fractionation (PFF) chip, where bubbles are sorted to size (Sample 3). (D) Schematic drawing of the acoustic characterization setup used to characterize the bubble populations by narrowband scattering and attenuation measurements. (E) Schematic drawing of the pulse-echo setup with a linear array transducer. The setup is used to study the nonlinear microbubble response in a typical focused ultrasound field employed for contrast-enhanced ultrasound imaging.  (F) The focal distances were dynamically varied from~2 to 7~cm in steps of 1~cm while the focal pressure was kept constant.

\item[Figure 2:]  Measured frequency dependent attenuation coefficients of (A) the native contrast agent, (B) the size sorted agent, and (C) the acoustically sorted agent. Attenuation curves were measured at peak negative pressures of 10, 25, 50, and 100~kPa. The simultaneously measured scattering coefficients at the fundamental requency are shown in Figs.~D-F and the second harmonic scattering coefficients are shown in Figs.~G-I. 

\item[Figure 3:]  Sensitivity expressed as the mean scattering cross-section per bubble $\overline \sigma_s$, of the acoustically sorted agent, of the size sorted agent, and of the native BR-14 agent plotted at (A) the fundamental frequency and (B) at the second harmonic as a function of the insonation pressure. 

\item[Figure 4:] (A) Scattered power at an ultrasound frequency of 1.5~MHz normalized by the power of the transmit pulse as a function of the imaging depth for axial focal distances of 2, 3, 4, 5, 6, and 7~cm for the (A) native agent and (C) for the acoustically sorted agent.  The white dots in (A) indicate the focal regions of the transmitted acoustic pressure fields. (B) Scattering response at the second harmonic plotted for the native agent and (D) for the acoustically sorted agent. (E) Attenuation of the acoustically sorted  agent at an ultrasound frequency of 1.5~MHz as a function of the acoustic insonation pressure. (F) Modeled scattering response of acoustically sorted BR-14 microbubbles.


\section*{Supplementary information}
\item[Figure S1] (A) Scattered power at an ultrasound frequency of 1.5~MHz normalized by the power of the transmit pulse as a function of the imaging depth for axial focal distances of 2, 3, 4, 5, 6, and 7~cm for the (A) native agent and (C) for the acoustically sorted agent. The bubbles were insonified by 4~cycle ultrasound pulses at a frequency of 1.5~MHz. (B) Scattering response at the second harmonic plotted for the native agent and (D) for the acoustically sorted agent.

\end{description}

\makeatletter
\renewcommand{\fnum@figure}{\figurename~S1}
\makeatother

\begin{figure*}[htb] 
\begin{center}
\includegraphics[width=1\textwidth]{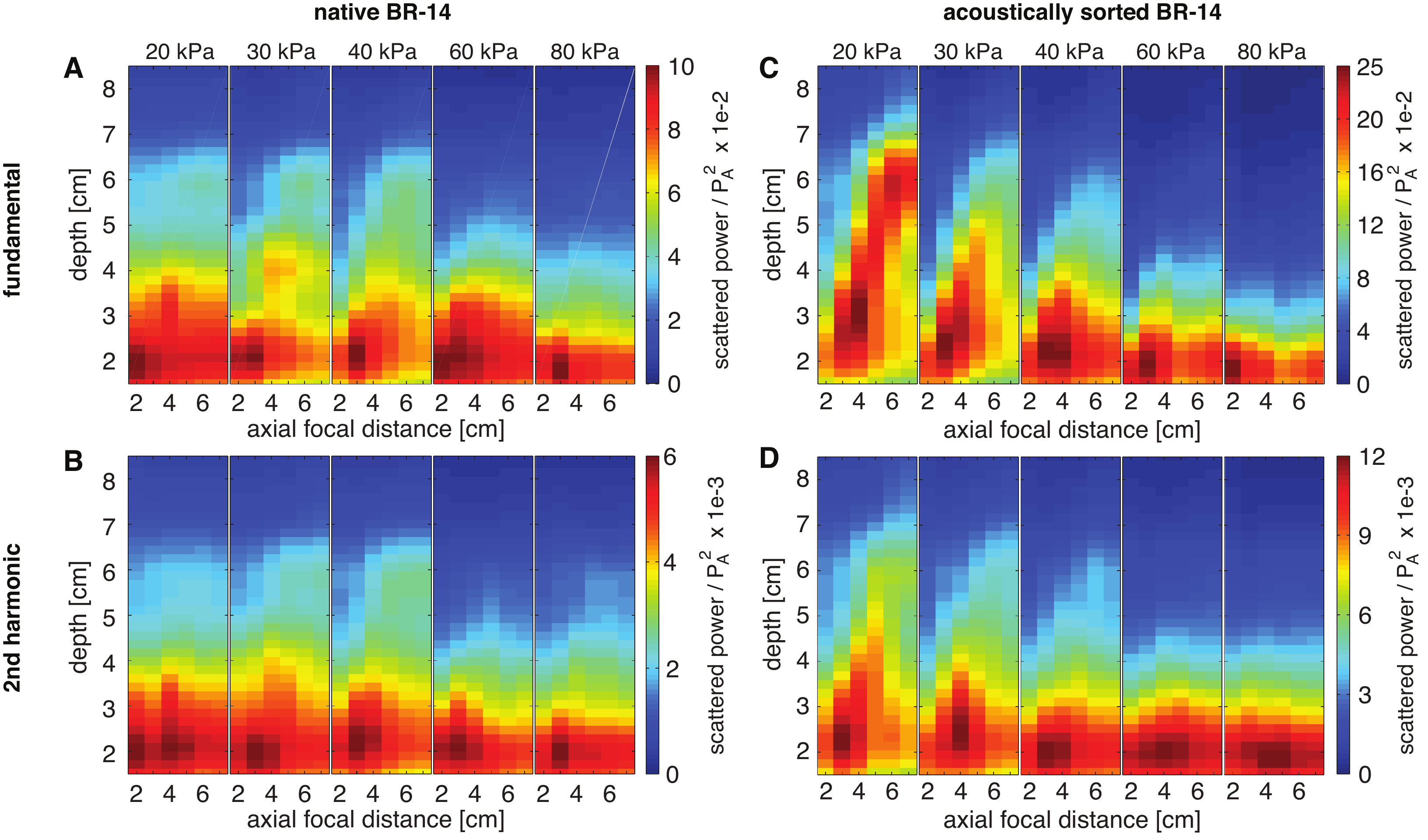}
\caption{ Pulse-echo response for  4~cycle, 1.5~MHz ultrasound transmit pulses. (A) Scattered power at an ultrasound frequency of 1.5~MHz normalized by the power of the transmit pulse as a function of the imaging depth for axial focal distances of 2, 3, 4, 5, 6, and 7~cm for the (A) native agent and (C) for the acoustically sorted agent.  (B) Scattering response at the second harmonic plotted for the native agent and (D) for the acoustically sorted agent.}
\label{F:S1}
\end{center}
\end{figure*}

\end{document}